\documentstyle[12pt] {article}
\begin {document}
\parindent=15pt
\begin{flushright}
{\bf US-FT/22-95}
\end{flushright}
\vskip .8 truecm
\begin{center}
{\bf A-DEPENDENCE OF HEAVY FLAVOUR PRODUCTION IN QCD}\\
\vskip 1.5 truecm
{\bf N.Armesto, C.Pajares, C.A.Salgado and Yu.M.Shabelski$^*$}\\
\vskip 0.9 truecm
{\it Departamento de Fisica de Part\'{\i}culas, Universidade de Santiago de
Compostela, \\
15706--Santiago de Compostela, Spain}
\end{center}
\vskip 2.5 truecm

\begin{abstract}
\hspace {30pt}

We calculate the A-dependence of charm and beauty production cross sections
on nuclear targets taking into account the difference of quark and gluon
distributions in free nucleons and in nucleus. At comparatively low energies,
if $\sigma \sim A^{\alpha}$, the value of $\alpha$ is slightly higher than
unity. With the growth of the initial energy the value of $\alpha$ decreases
and becomes smaller than unity. We also calculate the values of $\alpha$ for
different Feynman-$x$ of the produced $Q\overline{Q}$ pair and obtain that
they decrease significantly in the beam fragmentation region.

\end{abstract}
\vskip 2. truecm

*) Permanent address: Petersburg Nuclear Physics Institute, \\

Gatchina, Sanct-Petersburg 188350, Russia.
\vskip 1.5 truecm

August 1995 \\
\vskip .2 truecm

{\bf US-FT/22-95}

\pagebreak

\section {Introduction}
\vskip 0.5 truecm
The processes of heavy flavour production on nucleon and nucleus targets at
high energies are very interesting from both the theoretical and practical
points of view. These processes at high energies are usually considered in the
framework of perturbative QCD in the leading and the next-to-leading order
$\alpha_{s}$ expansion. In the case of a nuclear target it is usually assumed
that the cross section of  heavy flavour production, $\sigma(Q\overline{Q})$,
should be proportional to $A^\alpha$, with $\alpha = 1$, that is, in agreement
with the experimental result of [1], where
a value $\alpha = 1.02 \pm 0.03 \pm 0.02$ was obtained for the case of neutral
D-meson production at $\sqrt{s}$ = 39 GeV. However it is experimentally
well-known [2-8] that quark distributions in nucleus are slightly different
from the same distributions in free nucleons. So the value of $\alpha$ can
differ from unity and it seems interesting to estimate it. We will present in
the Section 2 the results obtained using the nucleus to nucleon structure
function ratios taken from Ref. [9].

\section {QCD predictions of charm and beauty production A-dependence}
\vskip 2mm

The standard QCD expression for heavy quark production cross section in a
hadron 1 - hadron 2 collision has the form

\begin{equation}
\sigma^{12\rightarrow Q\overline{Q}} = \int_{x_{a0}}^{1}
\frac{dx_a}{x_a}\int_{x_{b0}}^{1}
\frac{dx_b}{x_b}\left[x_aG_{a/1}(x_a,\mu^{2})
\right]\left[x_bG_{b/2}(x_b,\mu^{2})\right]
\hat{\sigma}^{ab\rightarrow Q\overline{Q}}(\hat{s},m_Q,\mu^{2}) \;,
\label{eq:totalpf}
\end{equation}
where $x_{a0} = \textstyle 4m_Q^2/ \textstyle s$ and
$x_{b0} = \textstyle 4m_Q^2/ \textstyle (sx_a)$. Here $G_{a/1}(x_a,\mu^{2})$
and $G_{b/2}(x_b,\mu^{2})$ are the \linebreak structure functions of partons
$a$ and $b$ inside hadrons $1$ and $2$ respectively, and \linebreak
$\hat{\sigma}^{ab\rightarrow Q\overline{Q}}(\hat{s},m_Q,\mu^{2})$ is the
cross section of the subprocess $ab\rightarrow Q\overline{Q}$ as given by
standard QCD. The latter depends on the parton center-of-mass energy
\mbox{$\hat{s} = (p_a+p_b)^2 = x_ax_bs$}, the mass of the produced heavy quark
$m_Q$ and the QCD scale $\mu^{2}$. Eq. (\ref{eq:totalpf}) should account
for all possible subprocesses $ab \rightarrow Q\overline{Q}$.

The parton cross section
$\hat{\sigma}^{ab\rightarrow Q\overline{Q}}(\hat{s},m_Q,\mu^{2})$ can be
written in the form \cite{kn:NDE}
\begin{equation}
\hat{\sigma}^{ab\rightarrow Q\overline{Q}}(\hat{s},m_Q,\mu^{2}) =
\frac{\alpha^{2}_{s}(\mu^{2})}{m^{2}_{Q}}f_{ab}(\rho ,\mu^{2}/m^{2}_{Q}),
\end{equation}
with
\begin{equation}
\rho = 4m^{2}_{Q}/\hat{s}
\end{equation}
and
\begin{equation}
f_{ab}(\rho ,\mu^{2}/m^{2}_{Q}) = f^{(0)}_{ab}(\rho) +
4\pi \alpha_{s}(\mu^{2})[f^{(1)}_{ab}(\rho) +
\hat{f}^{(1)}_{ab}(\rho)\ln{(\mu^{2}/m^{2}_{Q})}]\;.
\end{equation}

The functions $f^{(0)}_{ab}(\rho)$, $f^{(1)}_{ab}(\rho)$ and
$\hat{f}^{(1)}_{ab}(\rho)$ can be found in \cite{kn:NDE}.

For the numerical calculations we wrote the nuclear structure function
$G_{b/A}(x_b,\mu^{2})$ in the form
\begin{equation}
G_{b/A}(x_b,\mu^{2}) = A\cdot G_{b/N}(x_b,\mu^{2})\cdot R^A_b(x,\mu^{2}),
\end{equation}
similarly to Ref. \cite{ina} and we take the values of $R^A_b(x,\mu^{2})$
for gluons, valence and sea quarks from Ref. [9]\footnote{Let us note that the
theoretical estimations in an earlier paper \cite{kolya} give more or less a
similar behaviour of $R^A_b(x,\mu^{2})$ for light sea quarks but an enhancement
of gluon distributions in the nuclei at small $x$.}. They are presented in
Fig. 1. The values of $R^A_b(x,\mu^{2})$ in [9] are presented for $x > 10^{-3}$
that is not small enough at high energies. So at $x < 10^{-3}$ we used two
variants of the $R^A_b(x,\mu^{2})$ behaviour for gluon distributions: The first
is the extrapolation as $x^{\beta}$ with $\beta$ = 0.096 and 0.040 for the
charm and beauty production, respectively (solid lines in Fig. 1). Such
behaviour is in qualitative agreement with the results of \cite{lev}. The
second is the constant frozen at $x = 10^{-3}$ (dash-dotted lines). Three
different sets of parton distributions were used, namely MRS-1 \cite{mrs},
MT S-DIS \cite{mt} and GRV HO \cite{grv} that can be found in CERN PDFLIB
\cite{pdf}, but they give practically the same results for the $\alpha$
behaviour (see Table 1). We have used in the case of charm production the
values
$m_c$ = 1.5 GeV and $\mu^2$ = 4 GeV$^2$ and in the case of beauty production
$m_b$ = 5 GeV and $\mu^2$ = $m_b^2$.

The obtained results for $\alpha$ determined from the ratios of heavy quark
production cross section on a gold target and on the proton are presented in
the Table 1 for three sets of structure functions and frozen gluon
distribution ratios at $x = 10^{-3}$ and in Fig. 2 for the GRV HO set and two
variants of gluon distribution ratios in the small $x$ region. Here
$\sqrt{s_{NN}}$ is the c.m. energy for the interaction of the incident proton
with one target nucleon. One can see that at fixed target energies the values
of $\alpha$ are slightly higher than unity, which is not in contradiction with
the result of Ref. [1]. However $\alpha$ decreases with increasing energy and
this effect is larger in the case of charm production than in the case of
beauty. One can see also that the difference between two variants for
$R^A_b(x,\mu^{2})$ at $x < 10^{-3}$ becomes important only at the highest
energies.

We also calculate the values of $\alpha$ for different Feynman-$x$ ($x_F$)
regions using $x_F = x_a - x_b$ at energies $\sqrt{s_{NN}}$ = 39 GeV and
1800 GeV. The results are presented in Fig. 3. At negative and moderate $x_F$
(in the nucleus fragmentation region) the values of $\alpha$ are slightly
higher than unity. However in the case of charm production in the beam
fragmentation region (positive $x_F$) the values of $\alpha$ become essentially
smaller than unity. For beauty production the last effect is expected only
at very high energies.

\vskip 9 mm
\section {Conclusions}
\vskip 5 mm

We calculate the A-dependence of charm and beauty production using standard
QCD formulas and accounting in the difference of parton distributions for
free and bound nucleons. If one parametrize the heavy flavour production cross
section as $\sigma \sim A^{\alpha}$, the value of $\alpha$ is slightly
different from unity at the available energies. At comparatively low energies
the obtained values of $\alpha$ are a little larger than unity. This should be
connected with some nucleon-nucleon correlations which change the large-$x$
parton distributions.

At higher energies the values of $\alpha$ decrease and become smaller than
unity. At $\sqrt{s_{NN}}$ = 1800 GeV we expect the value of $\alpha \sim$
0.95. The decrease of the ratios $R^A_b(x,\mu^{2})$ that results in a decrease
of $\alpha$ can be connected with the effects of parton density saturation
\cite{glr} which occurs in heavy nuclei at $x$ values higher than in the
proton.

If we consider two small and different values of $x_a$ and $x_b$ in Eq. (1), it
is clear that the contribution to the inclusive cross section from the region
$x_a < x_b$ should be larger than the mirrow contribution ($x_a > x_b$) because
the value of the ratio $R^A_b(x,\mu^{2})$ in the first case is larger. It means
that heavy quark pairs will be produced preferably in the nucleus fragmentation
hemisphere, i.e., asymetrically, that is quite usual and confirmed
experimentally in the case of light quark hadron production. From Fig. 3 it is
clear that charm production on nucleus at LHC energies should give information
on the nuclear shadowing of the structure functions at small $x$.

Let us note also that at HERA-B energy ($\sqrt{s_{NN}}$ = 39 GeV) with GRV(HO)
parton distributions we obtain $\sigma(pp \rightarrow b\overline{b})$ = 6.8 nb
that is in agreement with the experimental value $5.7 \pm 1.5 \pm 1.3$ nb of
Ref. \cite{Jan}. However the experimental point should be decreased by 5\%
because it was obtained as an extrapolation from the gold target data assuming
an $A^1$ dependence whereas our calculations give the value $\alpha$ = 1.01.

We are grateful to M.A.Braun for useful discussions and both to K.J.Eskola and
to I.Sarcevic for sending us their numerical results. We thank the Direcci\'on
General de Pol\'{\i}tica Cient\'{\i}fica and the CICYT of Spain for financial
support. N.A. and C.A.S. also thank the Xunta de Galicia for financial support.
The paper was also supported in part by INTAS grant 93-0079.

\newpage

\begin{center}
{\bf Table 1}
\end{center}
\vspace{15pt}
Perturbative QCD predictions for the $\alpha$ values of charm and beauty
production in high energy $pA$ interactions with  gluon distribution ratios
frozen at $x = 10^{-3}$.
\begin{center}
\vskip 12pt
\begin{tabular}{|c||r|r|r|r|r|r|}\hline
  $\sqrt{s}, GeV$ & \multicolumn{2}{c|}{MSR-1} & \multicolumn{2}{c|}{MT S-DIS}
& \multicolumn{2}{c|}{GRV(HO)}    \\ \cline{2-7}
 & $c\overline{c}$ & $b\overline{b}$  & $c\overline{c}$ & $b\overline{b}$ &
 $c\overline{c}$ & $b\overline{b}$ \\
  \hline
  27      & 1.01 & 1.01 & 1.01 & 1.00 & 1.01 & 1.01     \\
  39      & 1.01 & 1.01 & 1.01 & 1.01 & 1.01 & 1.01   \\
  62      & 1.00 & 1.01 & 1.00 & 1.01 & 1.00 & 1.01    \\
  120     & 0.99 & 1.01 & 0.99 & 1.01 & 0.99 & 1.01    \\
  200     & 0.99 & 1.01 & 0.99 & 1.01 & 0.98 & 1.00     \\
  1800    & 0.96 & 0.99 & 0.96 & 0.99 & 0.95 & 0.99     \\
  14000   & 0.94 & 0.98 & 0.94 & 0.98 & 0.93 & 0.97    \\
\hline
\end{tabular}
\end{center}
\vspace{40pt}

\newpage

\begin{center}
{\bf Figure captions}
\end{center}
\vskip 0.5 truecm
\newcounter{eda}
%\begin{itemize}
\leftmargin 50mm
\begin{list}%
{Fig.~\arabic{eda} :}{\usecounter{eda}
\leftmargin 16mm \itemindent -9mm}
\vskip 0.1 truecm

\item The functions $R^A_G(x,\mu^{2})$, $R^A_V(x,\mu^{2})$ and
$R^A_S(x,\mu^{2})$ which determine the ratios of the distributions for protons
in the nucleus versus free protons for gluons (solid and dash-dotted curves,
see text), valence quarks (dashed curves) and sea quarks (dotted curves)
respectively for $\mu^2 =$ 4 GeV$^2$ (a) and $\mu^2 = m_b^2$ (b).

\item The energy dependence of $\alpha$ in the cases of charm and beauty
production for GRV HO structure functions and using extrapolated (solid curves)
and frozen at $x = 10^{-3}$ (dashed curves) ratios of gluon distributions.

\item The Feynman-$x$ dependence of $\alpha$ values in the cases of charm and
beauty production at $\sqrt{s_{NN}}$ = 39 GeV (a) and 1800 GeV (b) for GRV HO
structure functions and using extrapolated (solid curves) and frozen at
$x = 10^{-3}$ (dashed curves) ratios of gluon distributions.

\end{list}
\pagebreak

\end {document}
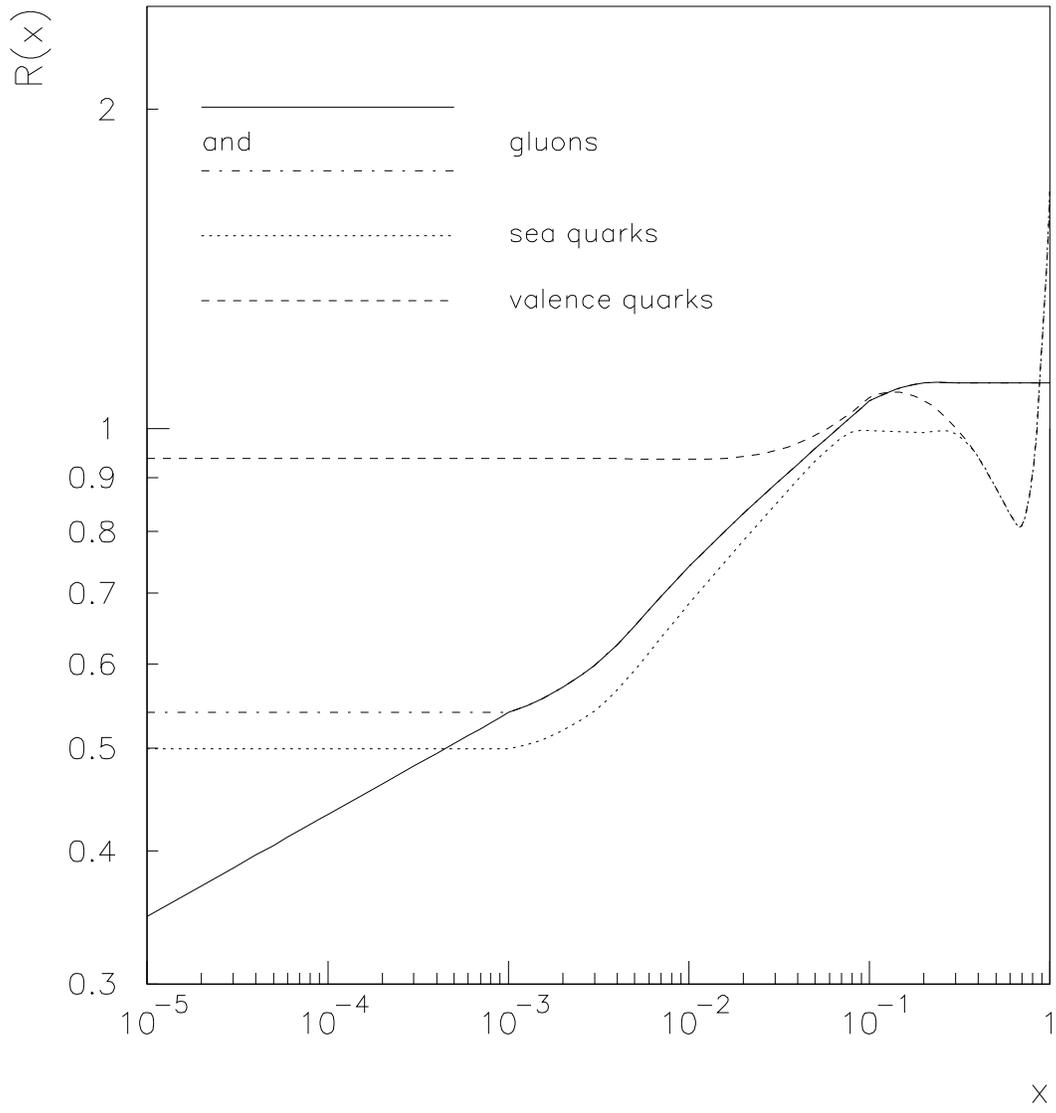
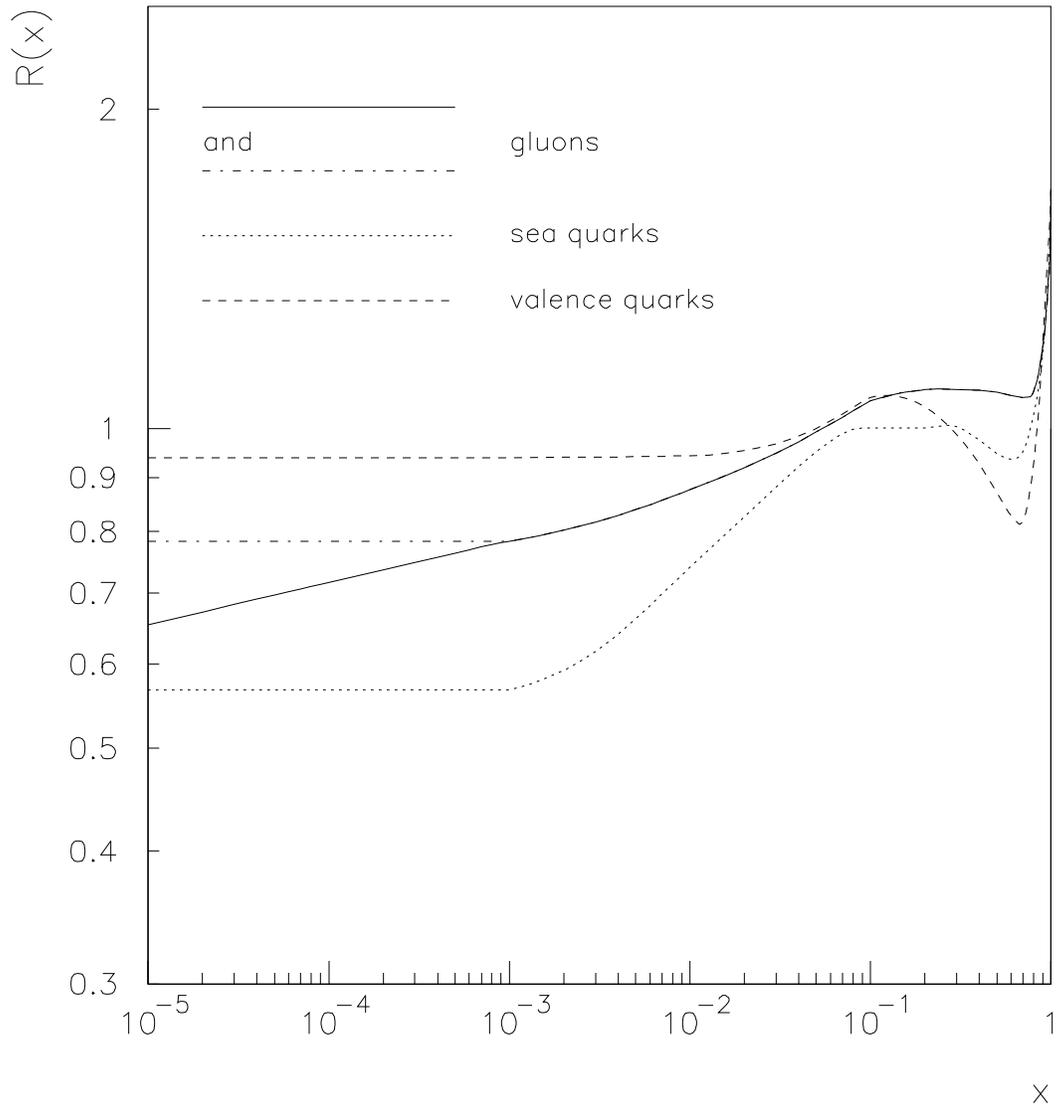
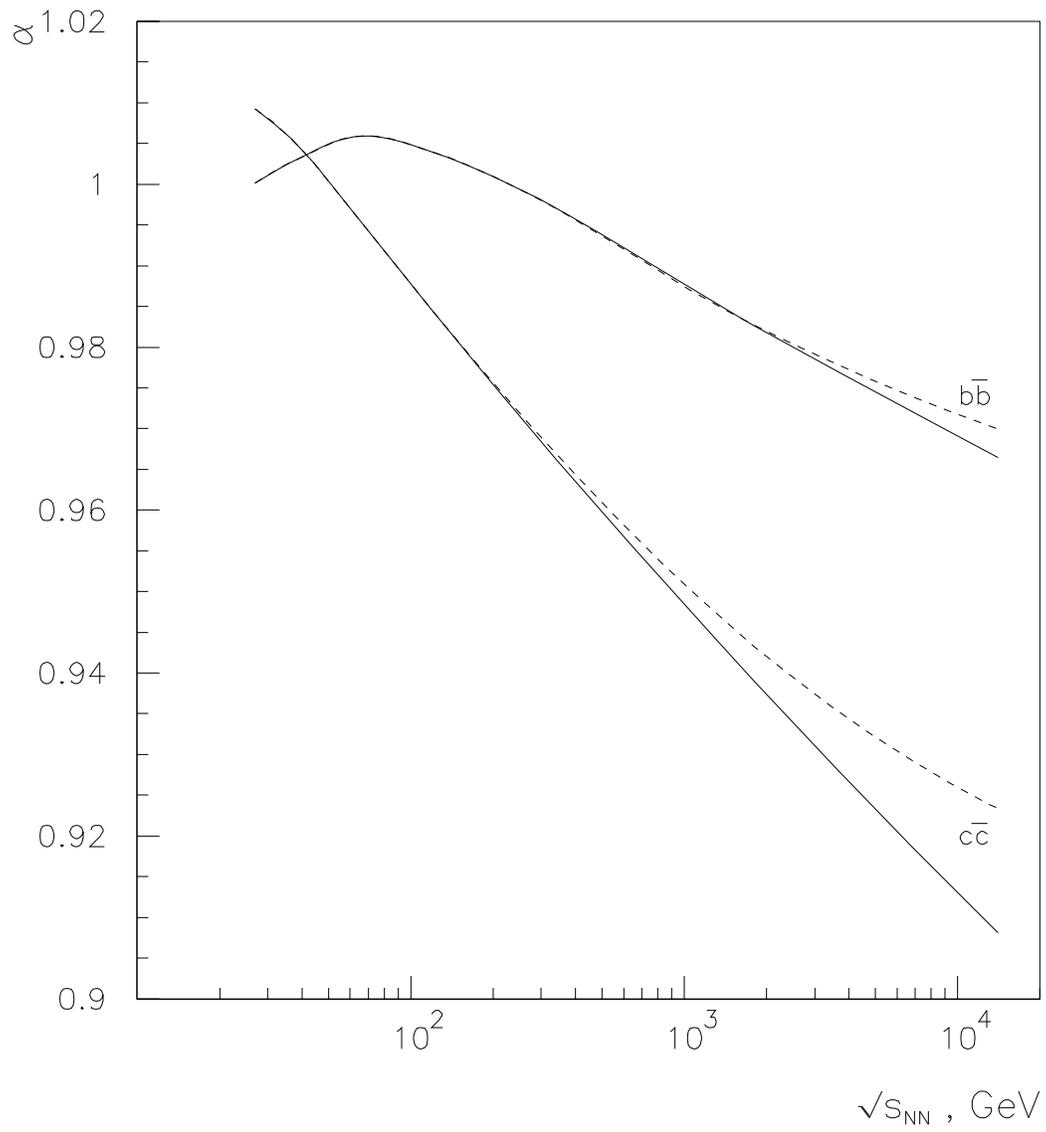
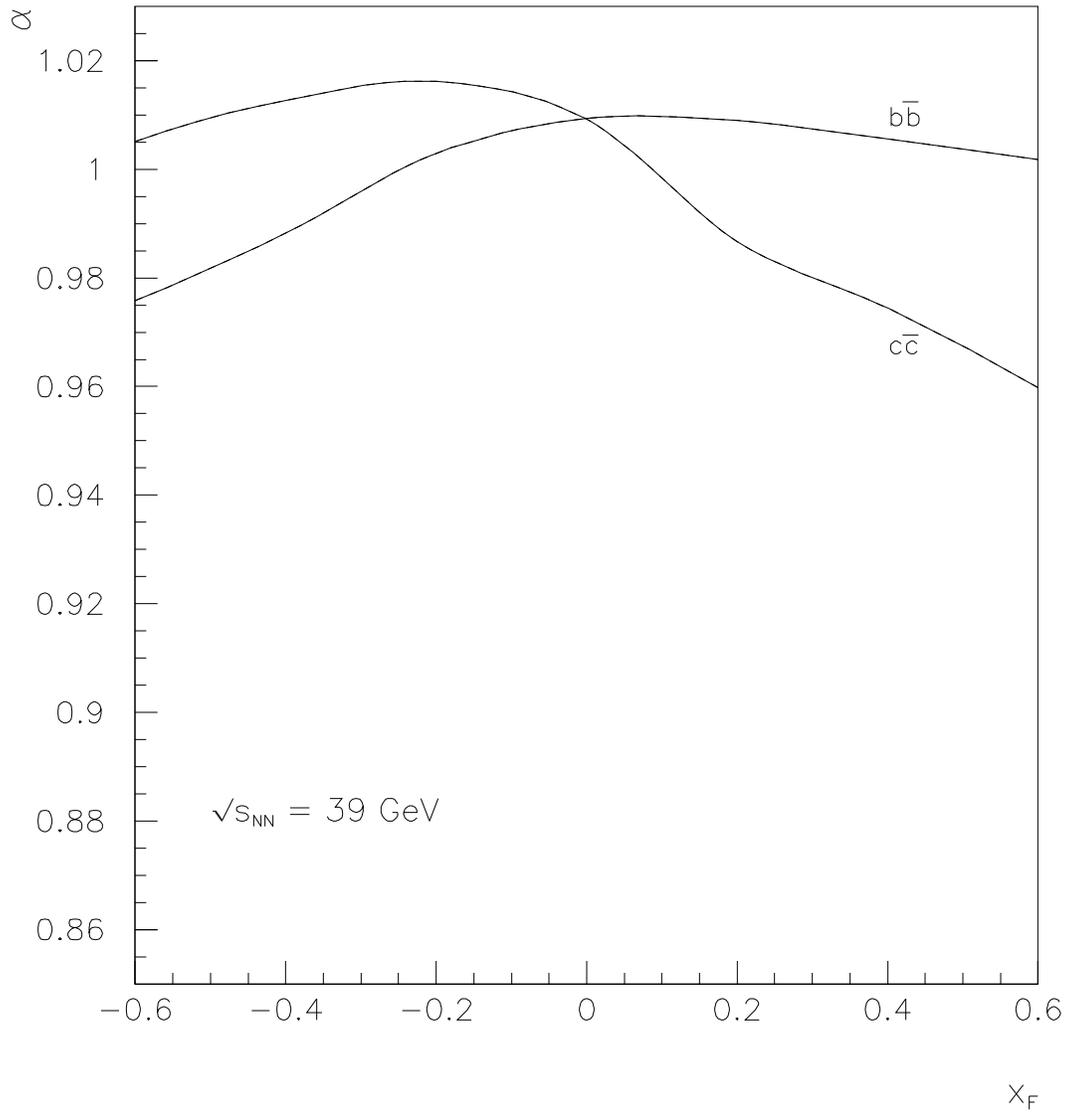
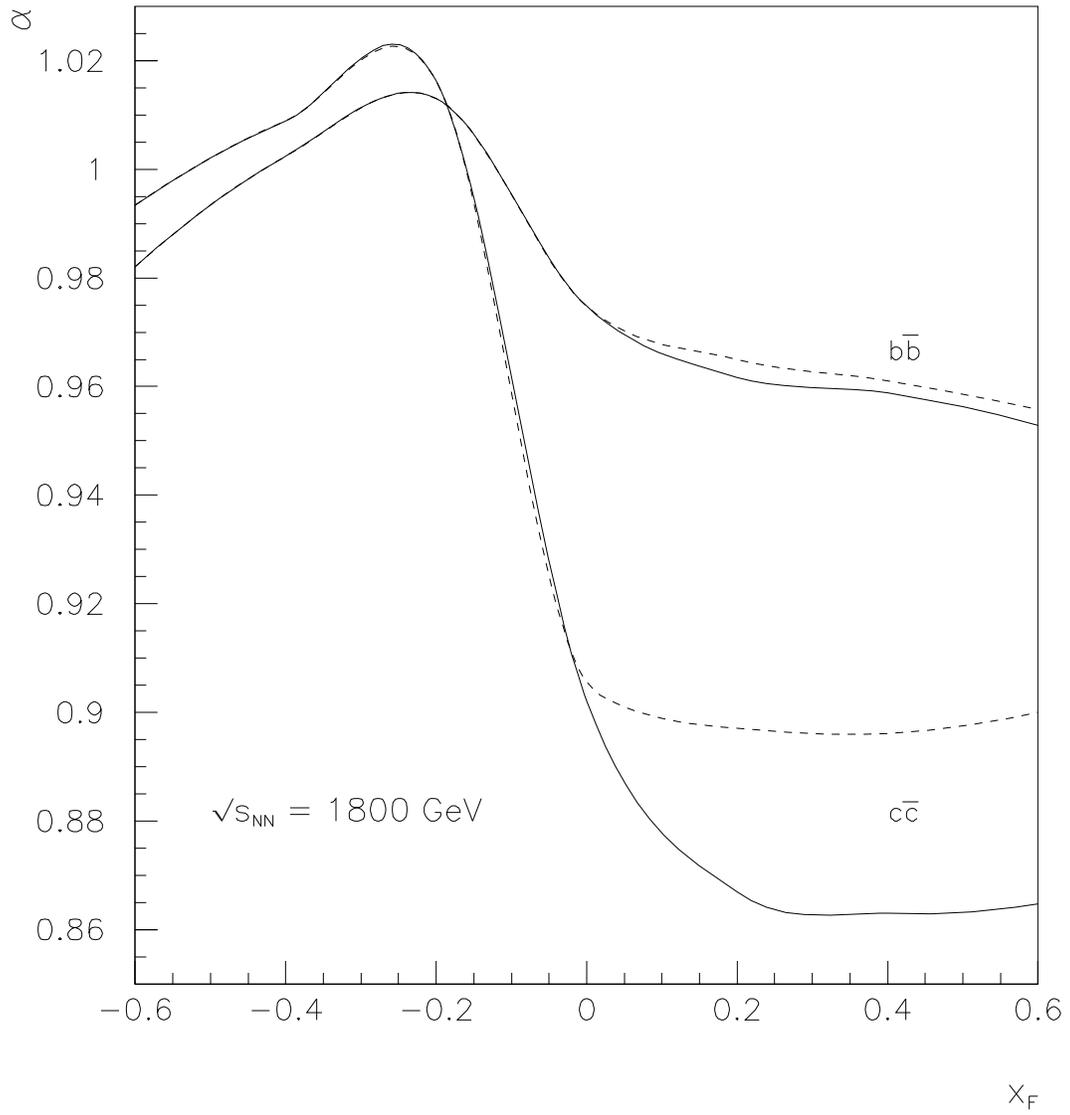
